\documentclass[sty/breakurl]{sty/eptcs}
\usepackage{cite}               
\usepackage{graphicx,latexsym,amssymb}               
\usepackage{amsmath}
\usepackage[parfill]{parskip}
\usepackage{color,sty/vmacros}   
\usepackage{hyperref}
\usepackage{xypic}

\def\Da{{\mit\Delta}}

\title{
Equilibrium and Termination
}
\author{
Vincent Danos \& Nicolas Oury
\institute{
School of Informatics\\University of Edinburgh
}
}

\def\titlerunning{Equilibrium and Termination}
\def\authorrunning{V.\@ Danos \& N.\@ Oury}

\begin{document}
\maketitle

\begin{abstract}
We present a reduction of the Turing halting problem (in the simplified form of the Post correspondence problem) to the problem of whether a continuous-time Markov chain (CTMC) presented as a set of Kappa graph-rewriting rules has an equilibrium. It follows that the problem of whether a computable CTMC is dissipative (ie does not have an equilibrium) is undecidable.
\end{abstract}

\section{Introduction}
In this note we explore an aspect of the relationship between the notion of equilibrium of a continuous time Markov chain (CTMC) and that of the traditional concept of termination in rewriting systems. Unlike in deterministic dynamical systems, a Markov chain equilibrium is not a definite state, but rather a probability over the state space which is invariant under the Markov semigroup, \emph{and} satisfies an additional property explained right below. 
%To discuss the nature of this stronger property and its physical interpretation we need a couple of definitions. 

Suppose given a CTMC with a matrix of rates $Q=(q_{ij})$ over a finite state space $I$, where $q_{ij}$ stands for the rate at which the chain jumps from $i$ to $j$. A  probability distribution $p$ on $I$ is said to be an \emph{equilibrium probability for $Q$} if for all $i$, $j$ in $I$:
\EQ{
\label{aaa}
p(i)\cdot q_{ij} = p(j)\cdot q_{ji}
} 
In plain words, this definition is saying that, at equilibrium, the probability of observing a jump between $i$ and $j$ is the same as that of observing a jump from $j$ to $i$. In some sense, time has disappeared. (The equilibrium property is called having \emph{detailed balance} in chemistry, and being \emph{reversible} in probability theory.) Note that for (\ref{aaa}) to have a solution, one needs $q_{ij}=0$ if and only if $q_{ji}=0$. When a solution $p$ exists (and the underlying transition system is strongly connected), $p$ is the unique steady state of the chain, meaning the chain converges to $p$ no matter where it starts. The converse is not true. It is possible that the chain has a steady state which does not satisfy (\ref{aaa}). 

The existence of an equilibrium is equivalent (at least in the finite case) to the existence of an \emph{energy function for $Q$}, by which we mean a real-valued function $E$ on $I$ such that for any related states $i$, and $j$, $\exp(-E(i))\cdot q_{ij}=\exp(-E(j))\cdot q_{ji}$. So, if equilibrium is the disappearance of time, exhibiting an energy function is analogous to finding a termination proof. And if one follows on the analogy, it should be possible to prove that the problem of finding such a energy function for sufficiently expressive languages of CTMCs is undecidable. This is what we do here. Specifically, we consider a class of stochastic graph rewriting systems, defined in the Kappa language, and prove that an instance $X$ of the Post correspondence problem (a simple recursively enumerable-complete problem due to Emil Post) has a solution if and only if a corresponding Kappa rule set $R_X$ is \emph{dissipative} (ie admits no energy function). Essentially $R_X$ performs a general enumeration of candidate solutions for $X$. The reversibility of the search steps guarantees that no stone is left unturned. 

The choice of the Post correspondence problem and the Kappa language makes the encoding rather simple and pleasing. (For the reader who would like to test the encoding, Kappa can be obtained at \url{kappalanguage.org}.) As Kappa is used in the modelling of combinatorial biological molecular networks (as one of the languages following the rule-based approach, see eg~\cite{Kuttler:2009p69,kuhn2009formal,KrivineDB09}), this undecidability result also presents an interesting first step, and it is hoped a valuable warning sign, in the study of thermodynamically consistent rule-based models of such networks. 

We start this note with a reminder of the notions of equilibrium and of the Post correspondence problem mentioned above, we then set up our encoding, and prove that it works. The conclusion will discuss further possibly intriguing consequences.

\section{CTMC equilibrium}
Suppose given a finite CTMC, that is to say, $I$ a finite state space, and $Q$ a rate matrix over $I$ (for a complete definition see~\cite{norris1998markov}). 

We write $q_{ij}$ for the rate from $i$ to $j$, and $G_Q$ for the transition graph on $I$ defined as $(i,j)\in{G_Q}$ if $q_{ij}>0$. We suppose that $Q$ is such that for any two states $i$ and $j$, if $q_{ij}>0$ then $q_{ji}>0$. In other words we suppose $G_Q$ is symmetric. As said, this is a necessary condition for the existence of an equilibrium.   We also write $t(e)$, $s(e)$ for the target and source of an edge $e$ in $G_Q$.

\DE
The equilibrium problem for $Q$ is to find a real-valued map $E$ on $I$ %\to \mbb (-\infty,+\infty]$ 
such that:
\EQ{
\label{eqeq}
\forall (i,j)\in G_Q:E(i)-E(j)=\ln(q_{ij}/q_{ji})
}
\ED
Such a function is called an \emph{{energy function for $Q$}}. It assigns, in particular, an energy difference $\Da E(i,j):=E(j)-E(i)=\ln(q_{ji}/q_{ij})$
to any pair of related states (equivalently to any edge in $G_Q$). Depending on $Q$, there might be no such function, or
there might be many (see below). 

Any map $E$ defines a probability on $I$ (note that we need infinite energy for $p(i)=0$):
\AR{
p(i)=\dfrac1Z\cdot{e^{-E(i)}}&\hbox{with }
Z=\sum_i e^{-E(i)}
}
The energy/probability correspondence is a bijection between energy maps and probabilities on $I$ - up to an additive constant for energy. 
Clearly, equation (\ref{eqeq}) is only a rephrasing of equation (\ref{aaa}).

Note that: 1) if $E$ is constant, $p$ is uniform; 2) according to the convention chosen here (which is the usual one), 
if $E(i)<E(j)$, or equivalently if ${\Da}E(i,j)>0$, then the equilibrium favours staying in $i$ over staying in $j$. That is to say the lower the energy, the more favoured the state at equilibrium.

In equation (\ref{eqeq}) we have $|I|$ unknowns, and as many indepedent equations as there are pairs of edges in $G_Q$ plus one (for normalising the distribution $p$) - so when do we have solutions? 
\PRO[Wegscheider]\label{W}
Problem (\ref{eqeq}) has a solution if and only if $\sum_{e\in\ga} \Da E(e)=0$ over every cyclic path $\ga$ on $G_Q$;
this solution is unique up to a choice of one additive constant per connected component 
in $G_Q$.
\ORP
Suppose $E$ is a solution, then for every path $\ga=e_1,\ldots,e_n$ on $G_Q$, one has
$\sum_{e\in \ga} \Da E(e) = E(t(e_n)) -E({s(e_0)})$.
This sum is zero if $s(e_0)=t(e_n)$, that is to say as soon as $\ga$ is a cycle (simple or not).
Conversely, suppose the condition holds. For each connected component $C$ of $G_Q$
pick a node $i_C$, and an arbitrary value $E_C$; for each $i\in C$ pick a path $\ga$ from 
$i_{C}$ to $i$ (this is possible because $G_Q$ is symmetric), and set:
\AR{
E(i)= E_{C}+\sum_{e\in\ga}\Da E(e)
}
by the condition, this does not depend on the choice of $\ga$, nor does it depend on the choice of $i_C$ (up to the choice of
another constant $E_C$). Clearly, it is a solution.\qed

This condition -due to Wegscheider~\cite{schuster1989}- will be referred to as the \emph{W-condition} in the sequel. 

\subsection{A simple Petri net example}
Let us examine an example which will give us an opportunity to 1) extend the above definitions to a countably infinite state space; 2) introduce a simple language of CTMCs that our language of choice, Kappa, will extend later. 

Consider a simple Petri net with two reversible transitions:
\AR{
\emptyset\leftrightarrow A\\
A\leftrightarrow B
}
The above defines a transition system with a countably infinite state space which can be described as the set of pairs $(n,m)$ where $n$ is the number of $A$s and $m$ the number of $B$s. To obtain a CTMC we have to define rates for each of the transitions. We assume that these rates are chosen in a way that the energy differences for creating an $A$ and a $B$ are respectively $E_1$ and $E_2$ (see also the diagram below). Let us check that this CTMC satisfies the W-condition. 

A cycle basis in the transition graph is formed by the following squares: 
\AR{
\xymatrix@M=8pt{
 {} n,m\ar[r]_{E_2}\ar[d]^{E_1}
&{} n-1,m+1\ar[d]_{E_1}\\
 {} n+1,m\ar[r]^{E_2}
&{} n,m+1
}
}
where both paths have the same energy differential $E_1+E_2$. Hence, the induced CTMC satisfies the W-condition (\ref{W}).
Specifically, if we set $E(0,0)=0$, we get $E(n,m)=(m+n)E_1+mE_2$ which defines the limit probability:
\AR{
p(n,m)=\dfrac1Z\cdot e^{-m(E_1+E_2)}\cdot e^{-nE_1}
}
There are two things worth noticing here. First, the limit probability does not depend on the rates of our pairs of transitions, but only on their ratio; this is expected. Second, the partition function $Z=\sum_{n,m} e^{-m(E_1+E_2)}e^{-nE_1}$ is bounded if and only if both $E_1>0$ and $E_1+E_2>0$. This is new. It means that because our state space is infinite - the W-condition is not enough to guarantee the existence of an equilibrium, and one has to add the above provisions. These are rather natural as they are saying that creating $A$s and $B$s, is energetically unfavourable.  If any of the conditions fail, the system creates an unbounded number of $A$s (if $E_1<0$) or $B$s (if $E_1+E_2<0$). 

Hence, thereafter, when we say that a countably infinite CTMC has an equilibrium, we mean that it satisfies the W-condition above, \emph{and}, that its partition function converges.

\section{PCP and the Kappa encoding}
The Post correspondence problem ({PCP}, or PC problem) is as follows. We are given a set $X$ of pairs of \emph{non-empty} words $(u_1,v_1)$, \dots, $(u_n,v_n)$ over some fixed alphabet $\Sig$, and we ask if there exists $p\geq1$ and $f:\ens{1,\ldots,p}\to\ens{1,\ldots,n}$, such that $u_{f(1)}\cdots u_{f(p)}=v_{f(1)}\cdots v_{f(p)}$.

As an example, consider the pairs 
$x_1=(aa,a)$,
$x_2=(ba,ab)$, and 
$x_3=(b,ab)$, then:
$$
x_1x_2x_3=(aa,a)(ba,ab)(b,ab)=(aabab,aabab)
$$ 
is a solution.
Simple as it is, the PC problem is undecidable if $\Sig$ has at least two symbols~\cite{post1946variant}.
 
The next thing we do is to encode this decision problem in the W-condition of a well-chosen Kappa system. 
We will suppose given an instance $X=\ens{(u_i,v_i); 0<i\leq n}$.

\subsection{Brief intro to Kappa}
To this effect, we need to briefly introduce Kappa~\cite{danos2004formal}. The language generalises that of Petri nets of which we have already seen an example in the previous section. One has various agent types each with a name and an associated finite set of sites. Sites can be used to bind other sites (with the restriction that any given site can be used at most once) and/or hold internal values ranging in a finite set. 

Here are the four types of agents that we will need for our encoding:
\\- a \emph{forward} agent  $F(s,i)$, and 
\\- a \emph{backward} one   $B(s,i)$, as well as 
\\- a \emph{symbol} agent   $S(l,r,x_\sig)$ where the site $x$ bears an internal state $\sig$ in $\Sigma+\ens{\ast}$, and 
\\- an \emph{index} agent   $I(l,r,x_i)$ where $x$ bears an internal state $i$ in $\ens{1,\ldots,n}+\ens{\ast}$.

The objects produced by combining agents are called \emph{site graphs}. One has specific rewriting rules that specify under which conditions agents bind, unbind, change internal states and get created or deleted. Rules have rates that determine uniquely a CTMC (usually countably infinite as for Petri nets) of which the states are site graphs, and of which the transitions are rule applications.

\begin{figure}[h!]
\centering
\includegraphics[width=200pt]{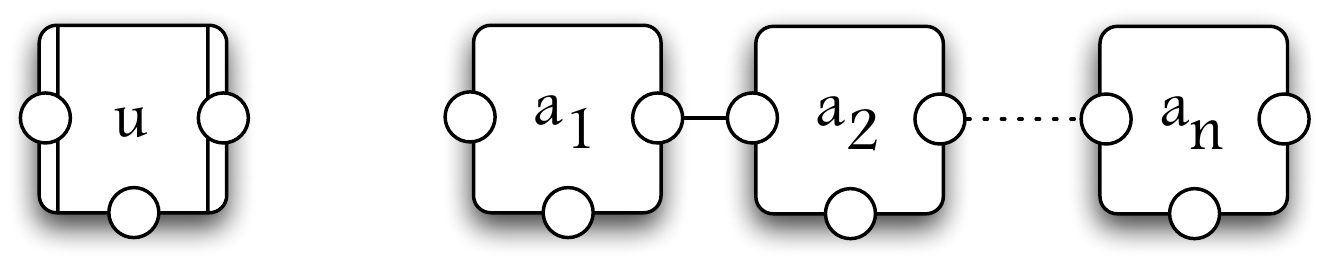} 
\caption{\small\it
The shorthand notation for $u=a_1\ldots a_n$ and its definition as an explicit chain of symbol agents. Note that the sites of the agents are understood from their position.}
\label{chain}
\end{figure}

All rewrite rules will be presented graphically as this is vastly more intuitive than 
textual syntax where links are presented as shared exponents, internal states as subscripts, 
and concatenated agents as separated by a comma. Eg for a chain of symbol agents representing
a word $a_1\ldots a_n$, we would have to write the cumbersome:
\AR{
S(l,r^1,x_{a_1}),S(l^1,r^2,x_{a_2}),\ldots,S(l^{n-1},r,x_{a_n})
}
where the connecting sites in the chain, $l$ and $r$ have names reminiscent of `left' and `right'. The actual integers used to identify connected sites are of no import. So instead, and equivalently, we shall use a graphical notation. Chains of symbol agents $u=a_1a_2\ldots a_n$ over $\Sig$ can be represented uniquely as indicated in Fig.~\ref{chain}, a notation which compares advantageously with the above one. We will elide the name of symbol and index agents, as it is easily recovered from context, and we will often do the same for site names, as they can be recovered unequivocally as well. This will permit a terse and visually pleasing presentation of the rule set encoding a PCP instance.

\subsection{Encoding}
\begin{figure}[h!]
\centering
\includegraphics[width=60pt]{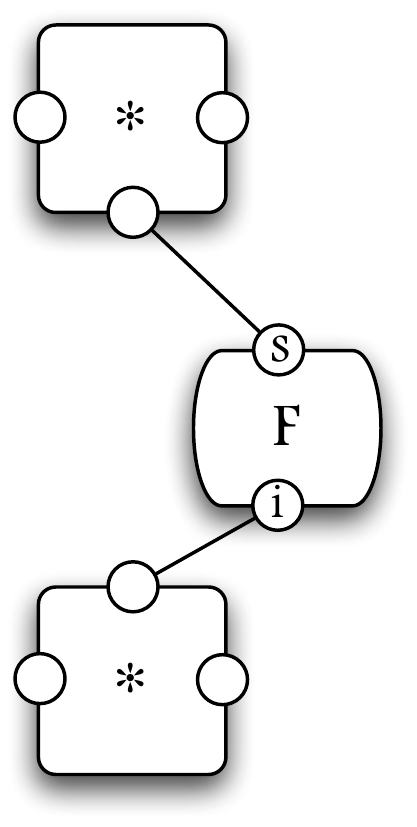} 
\caption{\small\it
The initial state of the $R_X$ system; the forward agent $F$ holds an empty symbol chain (upper dummy agent), as well as an empty index chain (lower dummy agent).}
\label{init}
\end{figure}

The idea of the encoding is that one starts in a state %$S(l,r,x^1_\ast),F(s^1,i^2),I(l,r,x^2_\ast)$ 
where the forward agent $F$ holds an empty word (on site $s$) and a dummy index (on site $i$) as shown in Fig.~\ref{init}. 
Both the dummy symbol and index are represented by $\ast$ (the internal state of $x$ is represented in the centre of each corresponding agent - purely for readability).

\begin{figure}[h!]
\centering
\includegraphics[width=270pt]{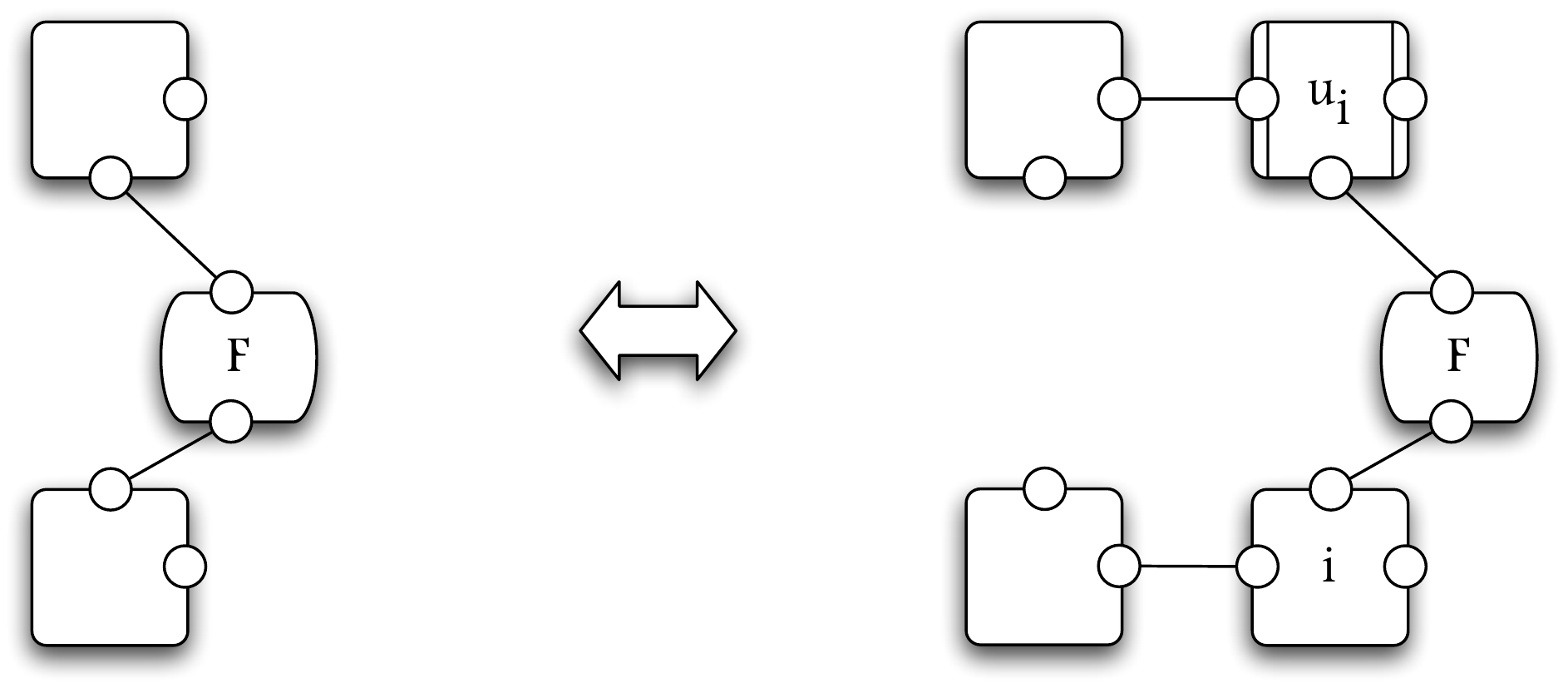} 
\caption{\small\it
$F$-rules in $R_X$ (with forward orientation from left to right): the forward agent $F$ extends the symbol chain by picking a word $u_i$ and records the index $i$ in the index chain.
Note that in the rule left hand side the current rightmost symbol and index agents need not be fully detailed (this basic Kappa convention is sometimes referred to as the `don't care, don't write' principle~\cite{KrivineDB09}). Note also that the shape $u_i$ is a
(non-empty) chain, not a single agent (as defined in Fig.~\ref{chain}).}
\label{Fui}
\end{figure}

As the computation proceeds, $F$ concatenates to the upper symbol chain new words picked in a non-deterministic fashion in $\ens{u_1,\ldots,u_n}$, while it records the index of these words in the lower index chain. The $n$ corresponding rules are depicted in Fig.~\ref{Fui}.

\begin{figure}[h!]
\centering
\includegraphics[width=210pt]{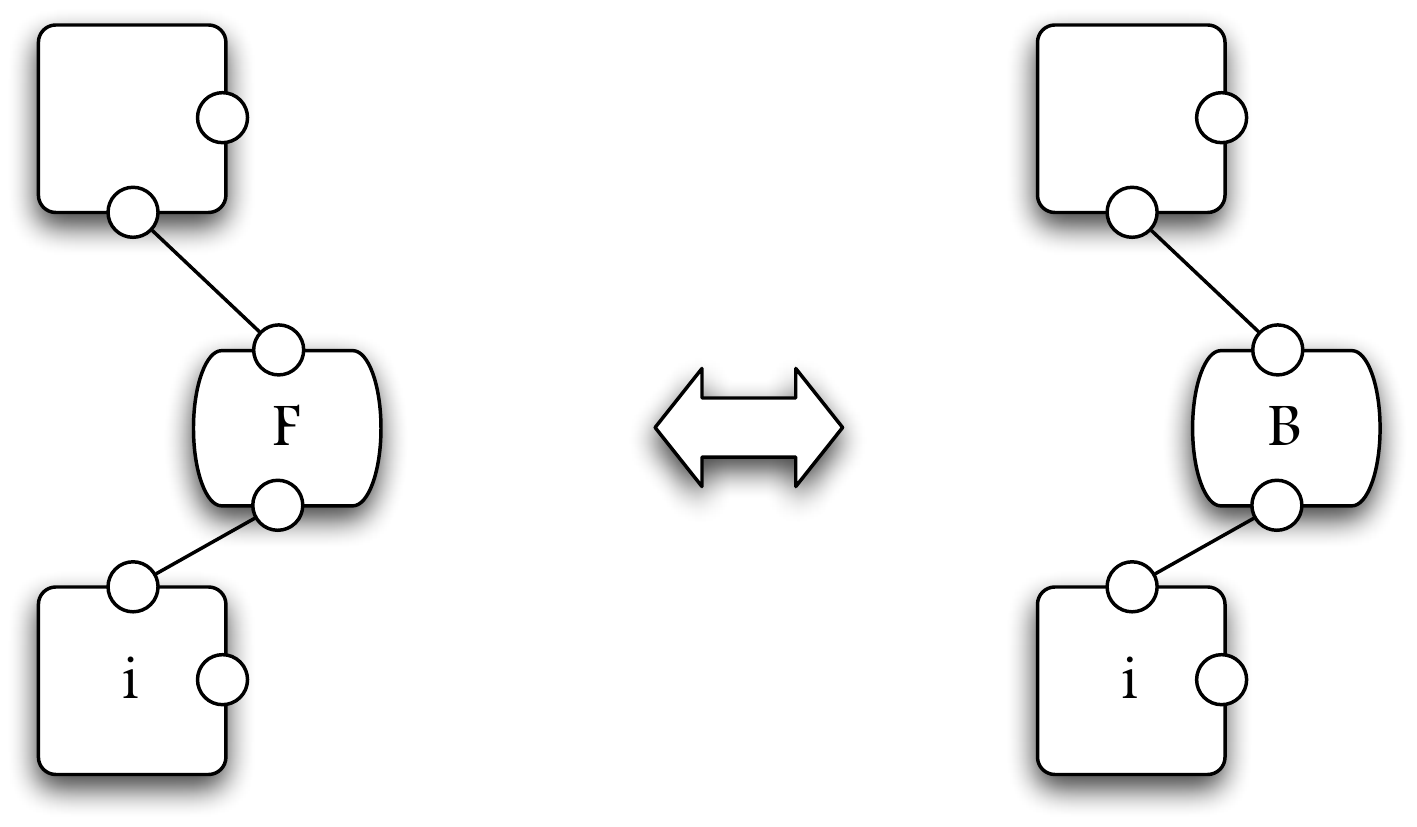} 
\caption{\small\it
Switching rules in $R_X$ (with forward orientation from left to right): under the condition that the $F$ agent has a non empty chain (the index one or equivalently the word one) it can flip into the $B$ agent.}
\label{switch}
\end{figure}

At some point $F$ will switch to a $B$ agent which will do the reverse work sliding down the index chain and re-parsing the symbol chain by chunking out words in $\ens{v_1,\ldots,v_n}$. Importantly, the switching rule(s) as shown in Fig.~\ref{switch} verifies that the index chain is not empty (ie the internal state of the lower index agent is a real index $i$, not a dummy one). This prevents the system to switch before anything has been done.

\begin{figure}[h!]
\centering
\includegraphics[width=280pt]{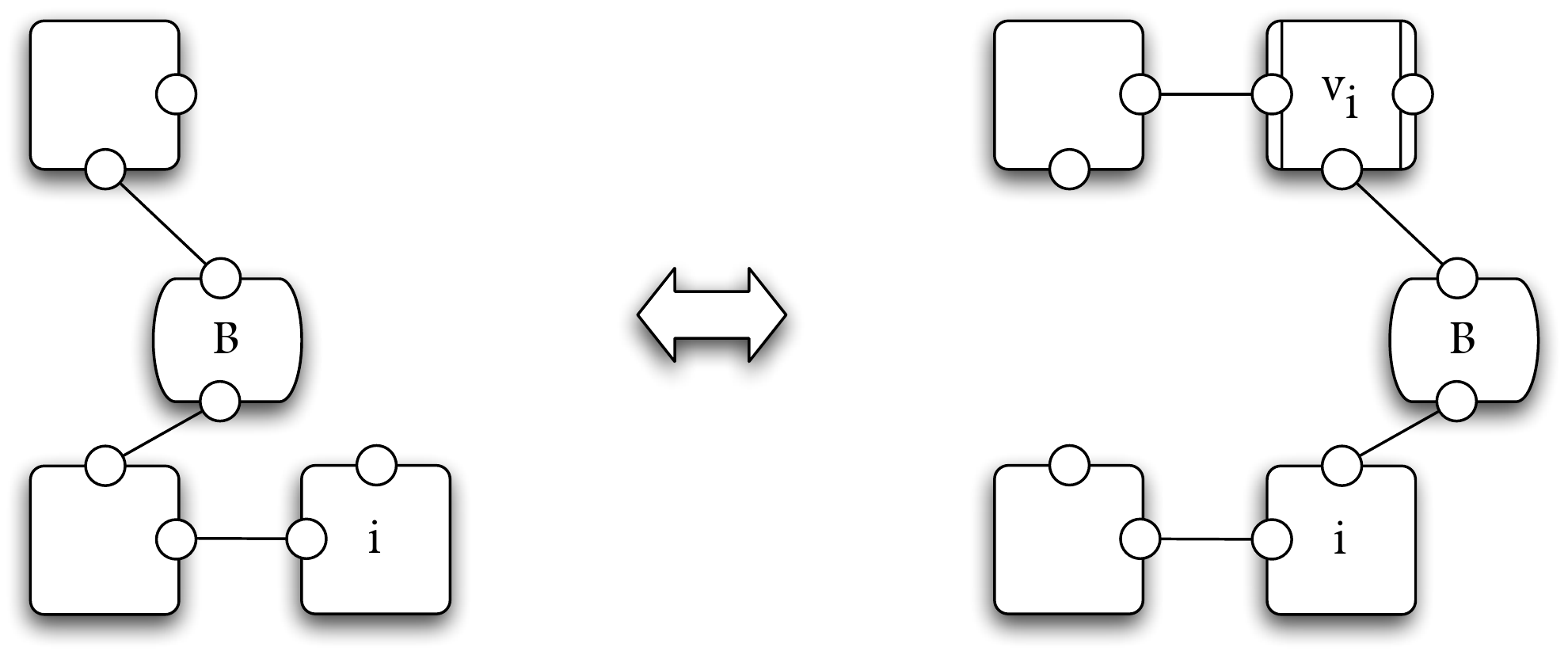} 
\caption{\small\it
$B$-rules in $R_X$ (with forward orientation from right to left): the backward agent $B$ re-parses the symbol chain backward by picking a word $v_i$ as indicated by the  index $i$ it is bound to in the index chain.
}
\label{Bvi}
\end{figure}

Once it has become a $B$, the middle agent slides backward on the index chain (which is a complete log of the choices made by the 
$F$ agent), and uses it to determine which next word to recognise. This is shown in Fig.~\ref{Bvi}. 

As indicated in Figs.\@ \ref{Fui}--\ref{Bvi}, each of the pairs of reversible rules we have considered so far has {a natural forward orientation}. Indeed, the agent $F$ natural orientation is to go \dots\ forward and extend the chains or eventually switch to the $B$ form, while the agent $B$ natural one is to consume the symbol chain and go backward. We refer henceforth to this natural direction as the \emph{{forward} direction}. By construction, going backward (aka backtracking) is deterministic, and any trace starting from the initial state can be visualised as an exploration of the PCP exploration tree, where backward steps allow backtracking and therefore guarantee at all times a path to any solution if there is one (possibly with infinite average hitting time). (This systematically available backtracking is reminiscent of the reversible CCS formalism of Ref.~\cite{danos2007formal}). In particular, agent $B$ can backtrack only by recreating the chain $v_i$ it has erased. It is the index chain/log that forces this. As a consequence, according to the rule set $R_X$ defined so far, $B$ cannot switch back to $F$ in any other state than the one at which $F$ itself switched.

From this, it is easy to see that \emph{success}, meaning a backward agent with an empty symbol chain, is equivalent to finding a solution to PCP. 

In fact, supposing all the rules in $R_X$ have non-zero rates we get:
\PRO
The set of solutions of a PCP instance $X$ is in bijection with the successful configurations
which the rule set $R_X$ can reach from the initial state. 
\ORP
As $B$ does not erase the index chain (see Fig.~\ref{Bvi}), a successful configuration (by definition one where agent $B$ has an empty symbol chain) contains a lower index chain which is a solution of the $X$ instance.

\subsection{Undecidability of the W-condition}
Now that we have a neat embedding of the search for solutions to $X$ as a rule set $R_X$, we need to relate its success to the equilibrium problem. The idea is the following. Because of the earlier remark on the deterministic nature of reverse steps, any rewrite trace is equivalent to a purely forward one, up to trivial cancellations. Thus, we are at liberty to add a series of new rules for $B$ to consume also gradually the index chain \emph{once} a success has been recorded. 

\begin{figure}[h!]
\centering
\includegraphics[width=310pt]{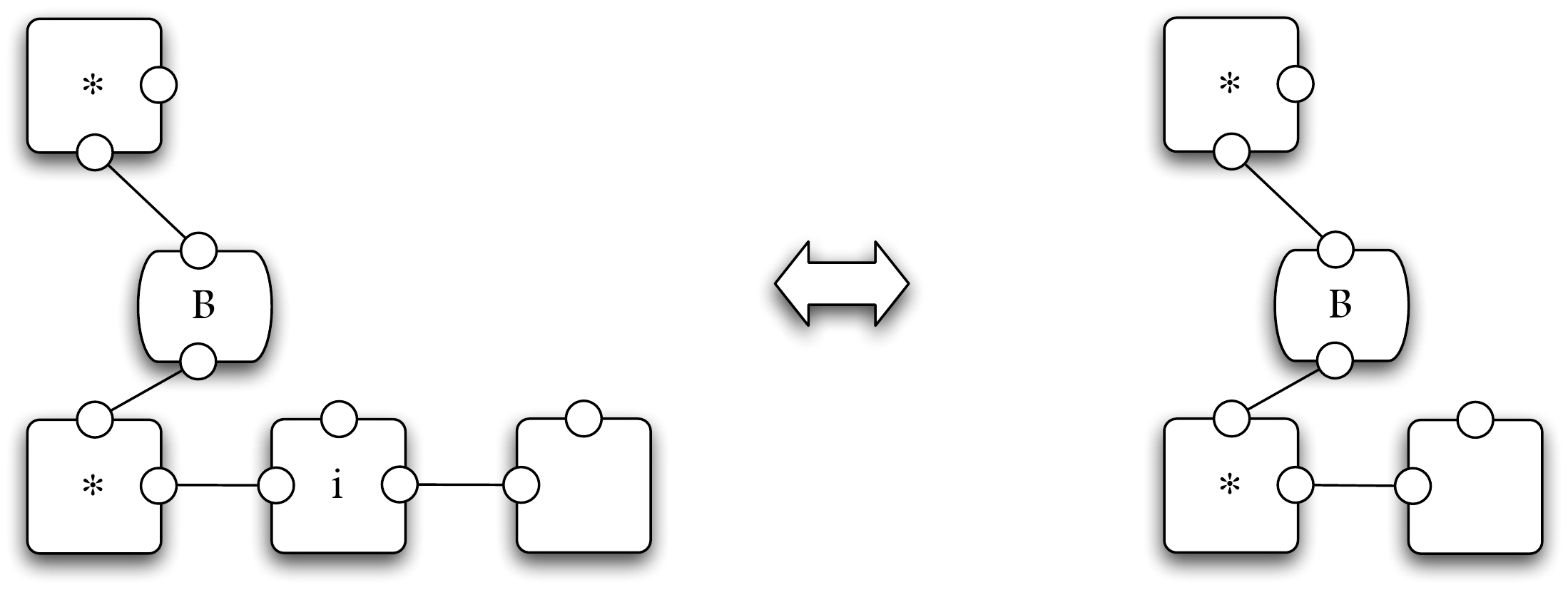} 
\caption{\small\it
Index chain deletions in $R'_X$ (with forward orientation from left to right): the $B$ agent, once a successful configuration is reached (as can be seen from the fact that the upper symbol chain is empty), progressively erases the index chain in order to return to the initial state.}
\label{swamp}
\end{figure}

\begin{figure}[h!]
\centering
\includegraphics[width=260pt]{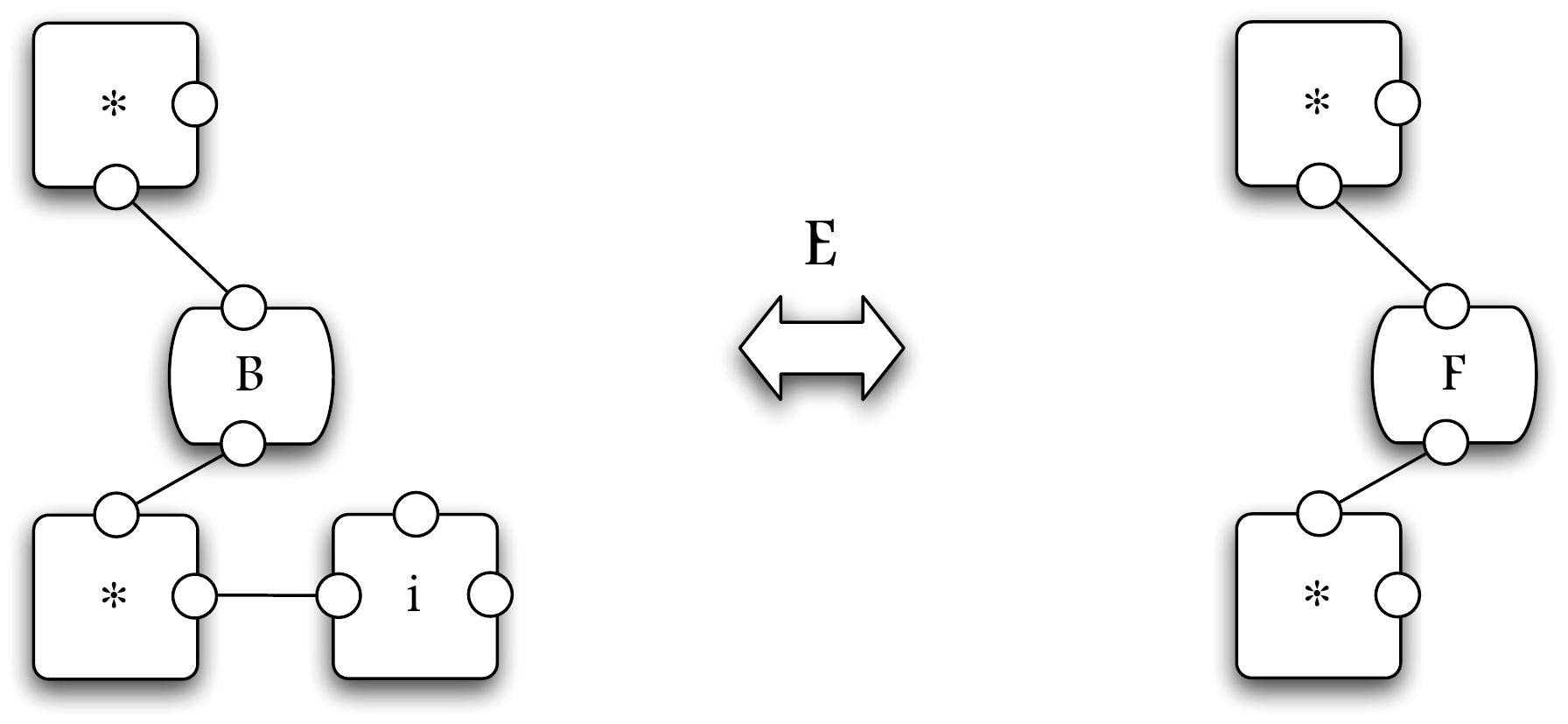} 
\caption{\small\it
Second switching rules in $R'_X$ (with forward orientation from left to right): the $B$ agent, once the index chain is erased (but for one remaining agent), flips into the $F$ agent, returning this to the initial state, and creating a loop conditioned on the existence of a success. The coefficient $E$ indicates the energy difference of the second switching rules (for the left-to-right direction).}
\label{swamp2}
\end{figure}

These new rules are shown
Fig.~\ref{swamp}--\ref{swamp2}. We call $R'_X$ the rule set formed by $R_X$ together with the new rules, and assign (non-zero) rates to all rules in $R'_X$ so as to obtain energy differences that are zero, except for the second switching rules of Fig.~\ref{swamp2}, where the energy difference is set to a constant $E\neq0$.

If $X$ has a solution, then one can simulate its discovery by a purely forward trace, which one can then conclude using the additional rules to return (in a forward way) to the initial state. This means there is a forward cycle in the state space. By construction its energy is $E\neq0$, which is a violation of the W-condition for $R'_X$, within the connected component of the initial state $C$ (thereafter called simply the initial component).

Conversely, suppose one has a violating cycle in $C$. The earlier remark about rewriting traces up to trivial forward/backward cancellations still applies with our bigger rule set $R'_X$. By definition, such cancellations do not change the energy difference associated to the path, nor the fact that it is a cycle. So we may assume our violating cycle $\ga$ has no such cancellations. For $\ga$ to violate the W-condition it must go through one of the second switching rules, as these are the only rules with a non zero energy differential. This means that one can take the origin of $\ga$ and choose its orientation in such a way that $\ga$ starts forward from the initial state and finishes with a second switching rule. But then $\ga$ must attain a successful configuration.

We have proved:
\PRO
A PCP instance $X$ has a solution if and only if the rule set $R'_X$ violates the W-condition.
\ORP

\subsection{Undecidability of dissipativity}
This proposition is not yet as strong as one would like. As we have seen earlier, for countably infinite state spaces, the W-condition is not enough to ensure an equilibrium, as the associated partition function might diverge. So we have not obtained yet the stronger result that $R'_X$ is dissipative (ie does not have an equilibrium) if and only if $X$ has a solution. 

Worse, with the particular chosen rates, this is clearly wrong. Let us see why. Reconsider the case where $X$ has no solutions. In this case the connected component $C$ of the initial state does not contain any success state, and therefore, from the point of view of $C$, the CTMC is entirely described by the rule set $R_X$. Since we have assigned a zero energy difference to all rules in $R_X$, every state in $C$ should have the same probability  (conditioned on the initial state being in $C$), and since $C$ is countably infinite, this is absurd and therefore can only mean that $Z$ diverges.

So, to get a convergent $Z$ we need to tweak our assignment of energies. It turns out that there is a very natural way to do this which is continuous with our previous construction. Pick a real number $\eps$, this will be our \emph{quantum of energy}. Assign to any state $x$ in $C$ the energy $n\cdot\eps$ where $n+1$ is the length of the index chain of state $x$ (equivalently $n$ is the number of non dummy indices in $x$'s index chain). 

Write $R_X'(\eps)$ for this more general assignment. The former energy assignment corresponds to $\eps=0$, ie $R'_X=R'_X(0)$. Except for the second switching rules, all transitions can be made compatible with this assignment (as stipulated in equation (\ref{eqeq})), as each induces a variation of the length of the index chain which is well defined. (For the second switching rules to be compatible, one would need $E=-\eps$.) So, clearly any cycle that does not use a second switching rule still has zero energy differential. 

As in the special case $\eps=0$, we reason that if there is a solution to $X$, there must be a $W$-violation, hence no equilibrium. Now, if there is no solution to $X$, the second switching rules are never used, else the $W$-condition is satisfied (by the point made just above). At this stage, we have recovered the $R'_X(0)$ argument. But this time we have more. 

\LE
Define $\Om_n$ as the set of states in the initial component $C$ with energy $n\cdot\eps$. If $X$ has no solution, then 
$|\Om_n|\leq(n+1)|X|^n$.
\EL
If $X$ has no solution, any state in $C$ with a given index chain of length $n+1$ is either a unique state with $F$ as the middle agent, or one of at most $n$ states with $B$ as the middle agent, depending on how far $B$ has slid back on the index chain. As there are $|X|^n$ such chains, the upper bound follows. \qed

Hence, if $\eps>\log|X|$, the associated partition function over $C$ converges:
\AR{
\sum_{x\in \Om_n} e^{-n\eps}\leq (n+1)|X|^ne^{-n\eps}\sim ne^{-n(\eps-log|X|)}
}
This implies:
\PRO
Let $X$ be a PCP instance, and suppose $\eps>\log|X|$, then $X$ has a solution if and only if the rule set $R'_X(\eps)$ is dissipative.
\ORP

Hence the problem of whether a countably infinite computable CTMC is dissipative is undecidable.

In passing, $\log|\Om_n|\approx n\log|X|$ is referred to in statistical physics as the \emph{entropy} of the energy equivalence class $\Om_n$ (note that the entropy is a macroscopic notion that presupposes a macro-observable - here the energy). So one can view our argument as saying that by fixing $\eps>0$ sufficiently positive, the entropy term can be controlled by the energy one. This is what physicists call a phase transition. In effect, all we need is to set a sufficient energy penalty on the exploratory behaviour of $F$ -that is to say its forward moves- to make the probing of longer potential solutions increasingly more expensive, and therefore more unlikely. In this argument, we have chosen a uniform penalty $\eps$ per increment of the index chain, but one could let $\eps$ depend on the chain length. (This leads to a probabilistic version of K\"onig's lemma, where the branching degree of the forward agent can be countered by a decreasing likelihood of exploring a branch.)

\section{Conclusion}
We have proved that the problem of whether a countably infinite computable CTMC is dissipative is undecidable. Early on (in \S2.1), we have described an example using a simple Petri net. Despite being complex, reachability is decidable for Petri nets and so they cannot host an encoding of PCP similar to the one we have used here. Nevertheless, there should be a refined version of the result that we have presented that would explain how difficult it is to determine whether a given Petri net is dissipative. One might speculate that this latter problem is at least as difficult as reachability. Likewise, it would be interesting to derive an NP version of our result. As bounded PCP (where the length of the solution is 
bounded at the outset) is NP-complete, one might think of using the same basic setting. This prompts another question. As said in the introduction, the PCP/Kappa couple does not play a fundamental role here. It is a way to get a precise formulation of the problem. It could be instructive to  attempt to repeat this argument at a more abstract level by using an axiomatic treatment of stochastic rewrite systems.

Another more practical research thread that is suggested here is directly related to the modelling issues at the heart of Kappa~\cite{KrivineDB09} and similar rule-based languages with a CTMC semantics such as the BNG one~\cite{blinov2004bionetgen,faeder2009rule}. It is the question of finding tractable forms of the W-condition that would be sufficient to ensure equilibrium (but obviously by our very result, not necessary). In the context of Kappa it is natural to think of introducing a class of energy functionals that would guarantee stronger and hopefully more feasible forms of the W-condition - perhaps based on the usage of local patterns as is customary in Ising models and derivatives thereof (eg see \cite[Chap.\@ 12]{pathria}). This is a problem of static analysis that we intend to investigate in the near future. 
Whichever structure one chooses to achieve this, it seems that in the context of model fitting, which is of cardinal importance in biological modelling (eg see Ref.~\cite{Gennemark:2009p145}), our result establishes that thermodynamic consistency has to be ``wired in'' the framework and can hardly be an afterthought.

\bibliographystyle{sty/eptcs}
\bibliography{bib/adhoc}
\end{document}